\documentclass[conference]{IEEEtran}
\usepackage{graphicx}
\usepackage{amssymb}
\usepackage{amsthm}
\usepackage{amsmath}
\usepackage{caption2}
\usepackage{subfigure}
\usepackage{mathtools}
\usepackage[utf8]{inputenc}
\usepackage{cite}
\ifCLASSINFOpdf
\else
\fi

\begin{document}
%
\title{Degrees-of-Freedom of the K-User MISO Interference Channel with Delayed Local CSIT}


\author{\IEEEauthorblockN{Chenxi Hao\authorrefmark{1} and Bruno Clerckx\authorrefmark{1}\authorrefmark{2}}
\IEEEauthorblockA{\authorrefmark{1}Department of Electrical and Electronic Engineering, Imperial College London,\\\authorrefmark{2}School of Electrical Engineering, Korea University.\\
Email: \{chenxi.hao10,b.clerckx\}@imperial.ac.uk}}

\maketitle

\IEEEpeerreviewmaketitle

\begin{abstract}
This paper considers a $K$-user Multiple-Input-Single-Output (MISO) Interference Channel (IC), where the channel state information obtained by the transmitters (CSIT) is perfect, but completely outdated. A Retrospective Interference Alignment (RIA) using such delayed CSIT was proposed by Maddah-Ali et. al for the MISO Broadcast Channel (BC), but the extension to the MISO IC is a non-trivial step as each transmitter only has the message intended for the corresponding user. Recently, Abdoli et.al focused on a Single-Input-Single-Output (SISO) IC and solved such bottleneck by inventing a distributed higher order symbol generation. Our main work is to extend Abdoli's work to the MISO case by integrating some features of Maddah-Ali's scheme. The achieved sum Degrees-of-Freedom (DoF) performance is asymptotically given by $\frac{64}{15}$ when $K{\to}{\infty}$, outperforming all the previously known results.\footnote{This work was supported in part by Samsung Electronics.}
\end{abstract}
\newtheorem{mytheorem}{Theorem}
\newtheorem{myremark}{Remark}
\section{Introduction}
Perfect and instantaneous knowledge of the CSIT on the interference link is crucial to the multiuser transmission in IC, but it is difficult to attain in practice. In a Frequency Division Duplex setup, each transmitter obtains the CSI of the corresponding user through feedback. This CSI is then shared among the transmitters via a backhaul link, so as to perform interference management. The latency incurred in this procedure is likely to be comparable with the channel coherence time. In the extreme case where the CSIT is completely outdated, the DoF performance achieved by Interference Alignment and conventional multiuser transmissions degrade to the case with no CSIT.

The usefulness of the completely stale CSIT was firstly found by Maddah-Ali and Tse in \cite{Tse10}, focusing on a $K$-user MISO BC. The invented scheme (known as MAT scheme) achieves the sum DoF $\frac{K}{1{+}\frac{1}{2}{+}{\cdots}{+}\frac{1}{K}}$, outperforming the case with no CSIT. As it will be clearer in Section \ref{sec:art_mot}, the transmission consists of $K$ phases, where in each phase, the transmitter 1) reconstructs the interferences overheard by the users in the previous phase using the perfect past CSIT; 2) multicasts them so as to provide additional useful signals for some users while align previously overhead interferences for some other users. This technique is termed as RIA later on.

The application to the $2$-user IC is reported in \cite{VVICdelay,xinping_mimo} by finding an alternative MAT scheme, which is distinct by the way of overheard interference retransmission. Nonetheless, such alternative design is only restricted to the two-user case. For the general $K$-user IC, since each transmitter only has access to the message intended for its corresponding user, the generalization of the $K$-phase transmission is a non-trivial step. Without the $K$-phase transmission, two $2$-phase schemes were reported in \cite{Ghasami11} and \cite{TorrellasK} for the MISO case, achieving the sum DoF $\frac{K^2}{K^2{-}K{+}1}$ and $\frac{2K}{K{+}1}$ respectively. For the 3-user SISO case, a 2-phase RIA scheme was proposed in \cite{MalekiRIA}, leading to a sum DoF of $\frac{9}{8}$. Recently, focusing on a $K$-user SISO IC, literature \cite{Abdoli13} invented a $K$-phase transmission, relying on a \emph{distributed higher order symbol generation} (to be discussed in detail in Section \ref{sec:art_mot}). Such a scheme improves the result to $\frac{36}{31}$ for $K{=}3$.

Thus far, to the best of our knowledge, there is no literature addressing the $K$-phase transmission for the $K$-user MISO IC with delayed CSIT. Hence, in this paper, we aim to extend the $K$-phase transmission proposed in \cite{Abdoli13} to the MISO case. Besides, the scheme integrates the higher order symbol transmission in the MAT scheme \cite{Tse10}. Another key ingredient is that new symbols transmission in phase 1 is performed via a proper transmitter scheduling. The achievable sum DoF is asymptotically given by $\frac{64}{15}{\approx}4{.}267$, significantly greater than $1$ in \cite{Ghasami11} and $2$ in \cite{TorrellasK}.

Organization: Section \ref{sec:system} elaborates the system model. Section \ref{sec:art_mot} revisits the prior art and highlights the motivation. The main results introducing the achievable scheme are presented in Section \ref{sec:dof}. Section \ref{sec:conclusion} concludes the paper.

Notations: Bold lower letters stand for vectors whereas a symbol not in bold font represents a scalar. $\left({\cdot}\right)^H$ denotes the Hermitian of a matrix or vector. ${\parallel}{\cdot}{\parallel}$ is the norm of a vector. $\mathbb{E}\left[{\cdot}\right]$ refers to the expectation of a random variable. $|\mathcal{S}|$ is the cardinality of the set $\mathcal{S}$. ${\lfloor}{a}{\rfloor}$ and ${\lceil}a{\rceil}$ stand for the greatest integer that is smaller than $a$ and the smallest integer that is greater than $a$, respectively.

\section{System Model}\label{sec:system}
In this contribution, we consider a MISO IC where there are $K$ transmitter-receiver pairs and the number of antennas at each Tx node is $K$ (but could also be greater than $K$). In a certain time slot $t$, denoting the transmitted signal from a certain Tx$k$ by $\mathbf{s}_k(t)$, of size $K{\times}1$, subject to the power constraint $\mathbb{E}[{\parallel}\mathbf{s}_k(t){\parallel}^2]{\leq}P$, the received signal $y_k(t)$ writes as
\begin{equation}
y_k(t){=}\sum_{j{=}1}^K\mathbf{h}_{kj}^H(t)\mathbf{s}_j(t){+}\epsilon_{k}(t).
\end{equation}
$\epsilon_k(t)$ represents the Additive White Gaussian Noise with zero mean and unit variance. $\mathbf{h}_{kj}$, of size ${K{\times}1}$, refers to the channel vector between Tx$j$ and Rx$k$. $\mathbf{h}_{kj}$ has circularly symmetric complex Gaussian entries with zero mean and unit variance (Rayleigh fading). The fading process is i.i.d across time slots (fast fading) and links.

We further assume that each receiver has perfect knowledge of the global CSI to perform the decoding. But the transmitters acquire their \emph{local} CSI with one-slot delay, due to the feedback mechanism and/or the backhaul link. As the channel is uncorrelated in time, the \emph{local} CSIT is completely outdated. In other words, at a certain time slot $t$, Tx$k$ perfectly knows $\{\mathbf{h}_{jk}(1){,}{\cdots}{,}\mathbf{h}_{jk}(t{-}1)\}{,}{\forall}j{=}1{,}{\cdots}{,}K$, while each receiver perfectly knows $\{\mathbf{h}_{ij}(1){,}{\cdots}{,}\mathbf{h}_{ij}(t)\}{,}{\forall}{i{,}j}{=}1{,}{\cdots}{,}K$.

A $K$-tuple rate $(R_1{,}{\cdots}{,}R_K)$ is achievable if each user decodes the desired message with arbitrary small error probability. Then, the system metric, i.e., sum DoF, is given by
\begin{equation}
d_s(K){=}\sum_{k{=}1}^Kd_k
{\triangleq}\lim_{P\to\infty}\frac{\sum_{k{=}1}^KR_k}{{\log}P}.\label{eq:sumdof}
\end{equation}


Moreover, for convenience, we reuse the same notation as in \cite{Abdoli13}, namely $u[i|\mathcal{S}_m;\mathcal{S}_{m^\prime}]$, to represent a symbol which is
\begin{itemize}
\item transmitted from Tx$i$ and made up of message intended for Rx$i$ only;
\item desired by a subset $\mathcal{S}_m$ of users, where $|\mathcal{S}_m|{=}m$;
\item already known by a subset $\mathcal{S}_{m^\prime}$ of users, where $|\mathcal{S}_{m^\prime}|{=}m^\prime$,
\end{itemize}
where $\mathcal{S}_m{\cap}\mathcal{S}_{m^\prime}{=}\emptyset$. With such a notation, we introduce two classes of symbols:
\begin{itemize}
\item Order-$m$ symbols, denoted by $u[i|\mathcal{S}_m]$ (i.e., $\mathcal{S}_{m^\prime}{=}\emptyset$), which is desired by a subset $\mathcal{S}_m$ of users, and known by no user;
\item Order-$(1{,}m^\prime)$ symbols, denoted by $u[i|i;\mathcal{S}_{m^\prime}]$ (i.e., $\mathcal{S}_{m}{=}i$, $|\mathcal{S}_m|{=}1$), which is intended for one Rx (i.e., Rx$i$), but already known by other $m^\prime$ users.
\end{itemize}

\section{Prior Art and Motivation}\label{sec:art_mot}
The MAT scheme proposed in \cite{Tse10} for MISO BC gives a fundamental idea of how to make use of the delayed CSIT. It has the following essential ingredients:

\underline{\emph{RIA}}: Let us consider a two-user case as an example, where the transmission lasts for two phases. All the new symbols are sent in phase 1, but decoding cannot be performed as each user does not have enough observations of the desired signal and also overhears interference. Essentially, in phase 2, using the perfect past CSI, the transmitter is able to reconstruct the overheard interferences and utilize them to formulate order-$2$ symbols. By multicasting those order-$2$ symbols, the symbols sent in phase 1 are decodable as each order-$2$ symbol provides an independent piece of useful signal for one user and allows for interference alignment for the other user.

\underline{\emph{$K$-Phase Transmission}}: Clearly, the number of order-$2$ symbols increases with the number of users in the system. Multicasting them one by one is time-consuming. Hence, for the case $K{\geq}3$, the transmission relies on a $K$-phase process, which achieves the optimal sum DoF performance in MISO BC. In a certain phase $m$, $K{-}m{+}1$ different order-$m$ symbols are delivered to a certain group of $m$ users and overheard by the other $K{-}m$ users. If these $K{-}m$ observations are provided to each of the scheduled users, the $K{-}m{+}1$ order-$m$ symbols become decodable thanks to the linear independent side information across the users. Towards this, at the beginning of the next phase, the transmitter reconstructs all the overheard interferences and utilizes them to generate order-$(m{+}1)$ symbols. The procedure ends till phase $K$ where no more higher-order symbols is generated.

However, extending the aforementioned techniques to the $K$-user MISO ($K{\geq}3$) IC is a non-trivial problem. In MAT, a certain order-$m$ symbol, which is generated for a certain set $\mathcal{S}_m$ of users, is made up of the messages intended for all the users in $\mathcal{S}_m$. But in IC, such higher order symbol cannot be obtained at any single transmitter because each transmitter only has access to the messages intended for its corresponding receiver. Due to this constraint, two attempts have been made in \cite{Ghasami11} (c.f. Theorem 5) and \cite{TorrellasK}, which consist of only two phases. All the new symbols are sent in the phase 1, generating order-$2$ symbols. In phase 2, the order-$2$ symbols are multicast one by one without generating higher order symbols ($m{\geq}3$). The achieved sum DoF is $\frac{K^2}{K^2{-}K{+}1}$ in \cite{Ghasami11} and $\frac{2K}{K{+}1}$ in \cite{TorrellasK}.

To overcome this bottleneck, Abdoli et.al in \cite{Abdoli13} have proposed a \emph{distributed higher order symbol generation} focusing on the SISO case. Specifically, at the beginning of phase $m{+}1$, using the perfect past CSI, Tx$k$ reconstructs the overheard interferences caused by its message, and utilizes them to formulate order-$(m{+}1)$ symbols (which are also made up of $c_k$). In other words, for a certain set of users $\mathcal{S}_{m{+}1}{=}\{i_1{,}{\cdots}{,}i_{m{+}1}\}$, there exists $m{+}1$ kinds of order-$(m{+}1)$ symbols intended for them, namely $u[i_1|\mathcal{S}_{m{+}1}]{,}u[i_2|\mathcal{S}_{m{+}1}]{,}{\cdots}{,}u[i_{m{+}1}|\mathcal{S}_{m{+}1}]$. Clearly, rather than being gathered by the single transmitter in MAT, those messages are partitioned in $m{+}1$ parts and to be sent from different transmitters. In addition, a new type of symbol, namely order-$(1{,}m)$ symbol, is generated in parallel. For $K{=}3$, such a scheme leads to a sum Dof of $\frac{36}{31}$, which is greater than $\frac{9}{8}$ that is achieved by a 2-phase scheme in \cite{MalekiRIA}. 

Motivated by \cite{Abdoli13}, an interesting work would be to investigate the benefit of such a distributed higher order symbol generation in the MISO case. To do so, we propose a scheme by integrating the higher order symbol transmission in the MAT scheme. Additionally, the proposed scheme performs interference management via transmitter scheduling, but in a different way than the state-of-the-art:
\begin{itemize}
\item \emph{Proper transmitter scheduling for fresh symbol transmission:} In phase 1, $n$ Tx/$n$ Rx are active, where $n{\leq}K$ is properly chosen based on the ability of higher order ($m{\geq}2$) symbol transmission, so as to boost the sum DoF performance. In \cite{Abdoli13,Ghasami11}, all Tx are active; in \cite{TorrellasK}, one transmitter is active per slot.
\item \emph{$1$ Tx/$m$ Rx scheduling in phase $m$ ($m{\geq}2$):} This not only ensures interference-free reception at the intended receivers, but also leads to enough linear independent overheard interferences at the unintended ones, in contrast to $2$ Tx/$m$ Rx scheduling in \cite{Abdoli13}, $1$ Tx/$2$ Rx in \cite{Ghasami11} (phase $2$), $2$ Tx/$2$ Rx in \cite{TorrellasK} (phase $2$). Note that there is no order-$m{,}m{\geq}3$ symbol transmissions in \cite{Ghasami11} and \cite{TorrellasK}.
\end{itemize}
%

\section{Main Results}\label{sec:dof}
\begin{figure}[t]
\renewcommand{\captionfont}{\small}
\captionstyle{center}
\centering
\includegraphics[width=0.36\textwidth,height=4.5cm]{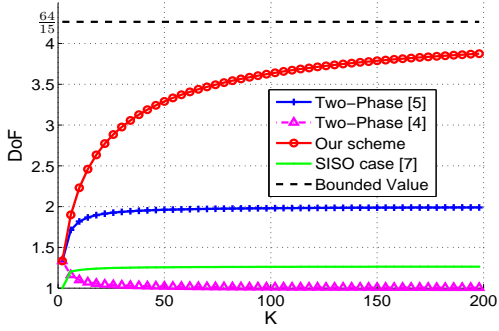}
\caption{Sum DoF performance}\label{fig:sumdof}
\end{figure}
\begin{mytheorem}\label{theo:DoF}
For a $K$-user MISO IC ($K{\geq}2$) where the transmitters are equipped with $K$ antennas, defining
\begin{equation}
\mathcal{O}(K){=}\left[1{-}\frac{1}{K{-}1}\sum_{l{=}2}^{K{-}1}\frac{K{-}l}{l^2{-}1}\right]^{-1},\label{eq:OK}
\end{equation}
and $\mathcal{O}_1(K){=}\lfloor2\mathcal{O}(K)\rfloor$ and $\mathcal{O}_2(K){=}\lceil2\mathcal{O}(K)\rceil$, the sum DoF $d_s(K)$ can be achieved, where
\begin{equation}
d_s(K){=}\max_{i{=}1{,}2}\frac{\mathcal{O}_i(K)^2}{1{+}\frac{\mathcal{O}_i(K)(\mathcal{O}_i(K){-}1)}{\mathcal{O}(K)}}.\label{eq:ds0}
\end{equation}
\end{mytheorem}
A comparison of Theorem \ref{theo:DoF} with the state of the art is shown in Figure \ref{fig:sumdof}. The sum DoF achieved by our scheme is bounded by $\frac{64}{15}$ (to be shown in the Appendix A). It converges much slower than all the other curves and dramatically outperforms the results in \cite{Ghasami11,TorrellasK,Abdoli13}. When $K{=}3{,}4{,}5$, one has $d_s(K){=}\frac{3}{2}$, $\frac{108}{65}$ and $\frac{360}{201}$ respectively. As it will be clearer later on, $\mathcal{O}(K)$ stands for the DoF of sending order-$2$ symbols, while $\mathcal{O}_{i^*}(K)$ (e.g. the optimal solution to \eqref{eq:ds0}) refers to the number of antennas employed in the new symbol transmission. If $\mathcal{O}_{i^*}(K){\leq}K{-}1$, Theorem \ref{theo:DoF} is also applicable to the case where each Tx is equipped with $K{-}1$ antennas ($K{\geq}2$).

Moreover, the only known upper-bounds in the context of the sum DoF with delayed CSIT are reported in \cite{Tse10} for K-user MISO BC and in \cite{VVICdelay,xinping_mimo} for two-user MIMO IC. These upper-bounds are obtained via a genie-aided model, which gives one user's observation to the others so as to construct physically degraded channels. However, in a K-user MISO IC, due to the reasons that 1) each transmitter only has the access to the message of its related user, and 2) each user overhears multiple interfers, the genie-aided model yields a loose upper-bound.

\subsection{Achievability Proof: $3$-User Case}\label{sec:example}
When $K{=}3$, one has $\mathcal{O}_1(3){=}2$ and $\mathcal{O}_2(3){=}3$, both of these values lead to $d_s(3){=}\frac{3}{2}$ according to \eqref{eq:ds0}. This implies that such sum DoF performance can be achieved by two approaches, which are distinct by the number of transmit antennas employed in phase 1. Let us firstly focus on the $3$-transmit antenna case. The $2$-transmit antenna case will be presented afterwards.

The sum DoF $\frac{3}{2}$ is achieved by sending $6$ symbols per Rx in $12$ slots. The transmission consists of three phases. In phase 1, $6$ symbols per Rx are transmitted in $2$ slots and $12$ order-$2$ symbols are generated. Phase 2 delivers those order-$2$ symbols in $6$ slots, resulting in $3$ order-$3$ symbols and $3$ order-$(1{,}2)$ symbols, which are transmitted using 3 slots in phase 3-I and 1 slot in phase 3-II respectively.
\subsubsection{Phase 1}\label{sec:M3phase1}
The transmission lasts for $2$ slots. In each slot, each Tx sends $3$ different symbols to its corresponding Rx. The transmitted signal writes as
\begin{equation}
\mathbf{s}_k(t){=}\mathbf{W}_k(t)\mathbf{x}_k(t), k{=}1{,}2{,}3,t{=}1{,}2\label{eq:sphase1M3}
\end{equation}
where $\mathbf{W}_k(t)$ is a full rank $3{\times}3$ precoding matrix and $\mathbf{x}_k(t)$ is $3{\times}1$ vector containing the private messages intended for Rx$k$. The received signal at a certain Rx$k$ is expressed as
\begin{multline}
y_k(t){=}\mathbf{h}_{kk}^H(t)\mathbf{W}_k(t)\mathbf{x}_k(t)\\{+}
\sum_{j{=}1{,}j{\neq}k}^3\underbrace{\mathbf{h}_{kj}^H(t)\mathbf{W}_j(t)\mathbf{x}_j(t)}_{u_t[j|j{,}k]}{,}k{=}1{,}2{,}3,t{=}1{,}2,
\label{eq:yphase1M3}
\end{multline}
where the noise term is ignored for convenience.

For clarity, let us focus on Rx1, who receives $\mathbf{x}_1(t)$ with other two interferences. Clearly, $\mathbf{x}_1(t)$ can be decoded if 1) $u_t[2|1{,}2]$ and $u_t[3|1{,}3]$ are removed; 2) $u_t[1|1{,}2]$ and $u_t[1|1{,}3]$ are provided to Rx1 in order to have enough linear independent observations of $\mathbf{x}_1(t)$. Similarly, Rx2 and Rx3 can decode their intended symbols if the interferences are removed and the useful side information is provided. In this way, $u_t[i|i{,}j]{,}i{\neq}j{,}$ is an order-$2$ symbol that is desired by Rx $i$ and $j$. Totally $12$ order-$2$ symbols result from these two slots. The remaining work consists in multicasting $u_1[1|1{,}2]$, $u_2[1|1{,}2]$, $u_1[2|1{,}2]$ and $u_2[2|1{,}2]$ to Rx1 and Rx2, $u_1[1|1{,}3]$, $u_2[1|1{,}3]$, $u_1[3|1{,}3]$ and $u_2[3|1{,}3]$ to Rx1 and Rx3 and $u_1[3|3{,}2]$, $u_2[3|3{,}2]$, $u_1[2|3{,}2]$ and $u_2[2|3{,}2]$ to Rx2 and Rx3.

Notably, in contrast to the SISO case \cite{Abdoli13} which employed a redundancy transmission (the number of slots is greater than the number of symbols intended for a single user) to create order-$2$ symbols, the above transmission generates order-$2$ symbol by making use of the linear independence of the i.i.d channel vectors across the users. As will be seen later on, the same philosophy is also applied to deliver the order-$m$ symbols ($2{\leq}m{\leq}K{-}1$).

\subsubsection{Phase 2}\label{sec:K3phase2}
\begin{table}[t]
\captionstyle{center} \centering
\renewcommand{\captionfont}{\small}
\begin{tabular}{c|cccc}
& Tx & Rx1 & Rx2 & Rx3\\ \hline
slot 5 & Tx2: $u_1[2|1{,}2]{,}u_2[2|1{,}2]$ & $y_1(5)$ & $y_2(5)$ & $u[2|1{,}2;3]$\\
slot 6 & Tx2: $u_1[2|3{,}2]{,}u_2[2|3{,}2]$ & $u[2|3{,}2;1]$ & $y_2(6)$ & $y_3(6)$\\
slot 7 & Tx3: $u_1[3|1{,}3]{,}u_2[3|1{,}3]$ & $y_1(7)$ & $u[3|1{,}3;2]$ & $y_3(7)$\\
slot 8 & Tx3: $u_1[3|2{,}3]{,}u_2[3|2{,}3]$ & $u[3|3{,}2;1]$ & $y_2(8)$ & $y_3(8)$
\end{tabular}
\caption{The transmission in slot 5 to 8 for the case $K{=}3$.}\label{tab:phaseIIM3}
\end{table}
The transmission is performed by transmitter scheduling. Specifically, each transmitter is active for two slots, multicasting two order-$2$ symbols for a certain subset of receivers in each slot. Meanwhile, the other two transmitters keep silent. The transmissions in slot 3 and 4 write as
\begin{IEEEeqnarray}{rcl}
\!\!\!\!\mathbf{s}_1(3)&{=}&\mathbf{W}_1(3)\left[u_1[1|1{,}2]{,}u_2[1|1{,}2]\right]^T,\IEEEyessubnumber\\
\!\!\!\!y_1(3)&{=}&\mathbf{h}_{11}^H(3)\mathbf{W}_1(3)\left[u_1[1|1{,}2]{,}u_2[1|1{,}2]\right]^T,\IEEEyessubnumber\\
\!\!\!\!y_2(3)&{=}&\mathbf{h}_{21}^H(3)\mathbf{W}_1(3)\left[u_1[1|1{,}2]{,}u_2[1|1{,}2]\right]^T,\IEEEyessubnumber\\
\!\!\!\!y_3(3)&{=}&\mathbf{h}_{31}^H(3)\mathbf{W}_1(3)\left[u_1[1|1{,}2]{,}u_2[1|1{,}2]\right]^T\!\!{=}u[1|1{,}2{;}3],\IEEEyessubnumber\\
\!\!\!\!\mathbf{s}_1(4)&{=}&\mathbf{W}_1(4)\left[u_1[1|1{,}3]{,}u_2[1|1{,}3]\right]^T,\IEEEyessubnumber\\
\!\!\!\!y_1(4)&{=}&\mathbf{h}_{11}^H(4)\mathbf{W}_1(4)\left[u_1[1|1{,}3]{,}u_2[1|1{,}3]\right]^T,\IEEEyessubnumber\\
\!\!\!\!y_2(4)&{=}&\mathbf{h}_{21}^H(4)\mathbf{W}_1(4)\left[u_1[1|1{,}3]{,}u_2[1|1{,}3]\right]^T\!\!{=}u[1|1{,}3{;}2],\IEEEyessubnumber\\
\!\!\!\!y_3(4)&{=}&\mathbf{h}_{31}^H(4)\mathbf{W}_1(4)\left[u_1[1|1{,}3]{,}u_2[1|1{,}3]\right]^T,\IEEEyessubnumber
\end{IEEEeqnarray}
where $\mathbf{W}_1(t){,}t{=}3{,}4$ is a full rank matrix of size $3{\times}2$. $u_1[1|1{,}2]$ and $u_2[1|1{,}2]$ (resp. $u_1[1|1{,}3]$ and $u_2[1|1{,}3]$) become decodable at Rx1 and Rx2 (resp. Rx3) if $u[1|1{,}2{;}3]$ (resp. $u[1|1{,}3{;}2]$) is provided to them as such a side information is linear independent of $y_1(3)$ and $y_2(3)$ (resp. $y_1(4)$ and $y_3(4)$).

Following the same framework, the transmissions in slot 5 to 8 are summarized in Table \ref{tab:phaseIIM3}. To sum up, the transmission is finalized if $u[1|1{,}2{;}3]$ and $u[2|1{,}2{;}3]$ are provided to Rx1 and Rx2, $u[1|1{,}3{;}2]$ and $u[3|1{,}3{;}2]$ are provided to Rx1 and Rx3, while $u[2|2{,}3{;}1]$ and $u[3|2{,}3{;}1]$ are provided to Rx2 and Rx3.
\subsubsection{Phase 3}\label{sec:K3phase3}
Following the distributed higher order symbol generation proposed in \cite{Abdoli13}, we form order-3 symbols as:
\begin{IEEEeqnarray}{rcl}
u[1|1{,}2{,}3]{=}&LC(u[1|1{,}2{;}3]{,}u[1|1{,}3{;}2]),\label{eq:order3K3}\IEEEyessubnumber\\
u[2|1{,}2{,}3]{=}&LC(u[2|1{,}2{;}3]{,}u[2|2{,}3{;}1]),\IEEEyessubnumber\\
u[3|1{,}2{,}3]{=}&LC(u[3|1{,}3{;}2]{,}u[3|2{,}3{;}1]),\IEEEyessubnumber
\end{IEEEeqnarray}
where $LC$ is short for \emph{Linear Combination}. $u[1|1{,}2{,}3]$, $u[2|1{,}2{,}3]$ and $u[3|1{,}2{,}3]$ are respectively transmitted using a single antenna by Tx1 in slot 9, Tx2 in slot 10 and Tx3 in slot 11 (namely phase 3-I). Consequently, Rx1 gets three independent interference-free observations of $u[1|1{,}2{;}3]$, $u[2|1{,}2{;}3]$, $u[2|1{,}2{;}3]$ and $u[3|1{,}3{;}2]$ as $u[2|2{,}3{;}1]$ and $u[3|2{,}3{;}1]$ can be removed by the past received signals at Rx1. The received signals at Rx2 and Rx3 follow similarly. So far, one more linear independent observation is needed to decode those four terms at each Rx. To this end, in the $12$th slot (phase 3-II), each Tx creates an order-$(1{,}2)$ symbol and transmits simultaneously. The order-$(1{,}2)$ symbols are given as
\begin{equation}
u[k|k;i{,}j]{=}LC(u[k|k{,}i{;}j]{,}u[k|k{,}j{;}i]){,}k{\neq}i{\neq}j,
\end{equation}
which should be linearly independent of \eqref{eq:order3K3} to prevent from aligning with the observations in phase 3-I. In this way, considering the received signals from slot 9 to 12, each Rx is able to decode the desired signal, so as to proceed to recover order-$2$ and private symbols.

Without the order-$3$ and order-$(1{,}2)$ symbols, the $12$ order-$2$ symbols created in phase 1 have to be delivered one by one, leading to the requirement of $12$ slots (rather than $10$). The sum DoF will be $\frac{18}{14}$, which is the same as in \cite{Ghasami11}. The new symbol transmission in the 2-phase scheme proposed in \cite{TorrellasK} works differently from our scheme. Although sending order-$2$ symbols sequentially yields the same sum DoF $\frac{3}{2}$ for $K{=}3$, it costs a huge number of time slots when $K$ is large.

Previous scheme relies on 3-transmit antennas in phase 1. Alternatively, we can also use 2-transmit antenna strategy in phase 1, which employs a transmitter-scheduling. Basically, every two transmitters are active per slot to deliver two new symbols to the corresponding receiver, and the pair of transmitters are scheduled in a round-robin fashion. Specifically, Tx1 and Tx2, Tx1 and Tx3, Tx2 and Tx3 are active in slot 1, 2 and 3 respectively, leading to the received signals as
\begin{IEEEeqnarray}{rcl}
y_1(1)&{=}&\mathbf{h}_{11}^H(1)\mathbf{W}_1(1)\mathbf{x}_1(1)
{+}\underbrace{\mathbf{h}_{12}^H(1)\mathbf{W}_2(1)\mathbf{x}_2(1)}_{u_1[2|1{,}2]},\IEEEyessubnumber\label{eq:y1t1M2}\\
y_2(1)&{=}&\underbrace{\mathbf{h}_{21}^H(1)\mathbf{W}_1(1)\mathbf{x}_1(1)}_{u_1[1|1{,}2]}
{+}\mathbf{h}_{22}^H(1)\mathbf{W}_2(1)\mathbf{x}_2(1),\IEEEyessubnumber\label{eq:y2t1M2}\\
y_1(2)&{=}&\mathbf{h}_{11}^H(2)\mathbf{W}_1(2)\mathbf{x}_1(2)
{+}\underbrace{\mathbf{h}_{13}^H(2)\mathbf{W}_3(2)\mathbf{x}_3(2)}_{u_1[3|1{,}3]},\IEEEyessubnumber\label{eq:y1t2M2}\\
y_3(2)&{=}&\underbrace{\mathbf{h}_{31}^H(2)\mathbf{W}_1(2)\mathbf{x}_1(2)}_{u_1[1|1{,}3]}
{+}\mathbf{h}_{33}^H(2)\mathbf{W}_3(2)\mathbf{x}_3(2),\IEEEyessubnumber\label{eq:y3t2M2}\\
y_2(3)&{=}&\mathbf{h}_{22}^H(3)\mathbf{W}_2(3)\mathbf{x}_2(3)
{+}\underbrace{\mathbf{h}_{23}^H(3)\mathbf{W}_3(3)\mathbf{x}_3(3)}_{u_1[3|2{,}3]},\IEEEyessubnumber\label{eq:y2t3M2}\\
y_3(3)&{=}&\underbrace{\mathbf{h}_{32}^H(3)\mathbf{W}_2(3)\mathbf{x}_2(3)}_{u_1[2|2{,}3]}
{+}\mathbf{h}_{33}^H(3)\mathbf{W}_3(3)\mathbf{x}_3(3).\IEEEyessubnumber\label{eq:y3t2M2}
\end{IEEEeqnarray}
where $\mathbf{W}_k(t)$ is a full rank $3{\times}2$ matrix, the symbol vector $\mathbf{x}_k(t)$ is of size $2{\times}1$. The received signals $y_3(1)$, $y_2(2)$ and $y_1(3)$ are not shown as Rx3, 2 and 1 keep silent in slot 1, 2 and 3 respectively. Similar transmissions are performed in slot 4, 5 and 6. Thus, totally $24$ new symbols (e.g. $8$ per Rx) are sent in $6$ slots, generating $12$ order-$2$ symbols.

Applying the higher order symbol transmission introduced in Section \ref{sec:K3phase2} and \ref{sec:K3phase3}, those $12$ order-$2$ symbols are successfully delivered in $10$ slots, yielding the sum DoF $\frac{24}{16}{=}\frac{3}{2}$. Clearly, this scheme is applicable to the case where each transmitter is equipped with $2$ antennas, as the transmission only relies on $2$ transmit antenna.

Remarkably, in the $3$-transmit antennas scheme, the decoding of every $3$ private symbols requires $4$ order-$2$ symbols, while in the $2$-transmit antennas scheme, the decoding of every $2$ symbols needs $2$ order-$2$ symbols, implying a more effective usage of order-$2$ symbols.

\subsection{Achievability Proof: Generalized Scheme}\label{sec:scheme}
\begin{figure}[t]
\renewcommand{\captionfont}{\small}
\captionstyle{center}
\centering
\includegraphics[width=0.5\textwidth,height=1.5cm]{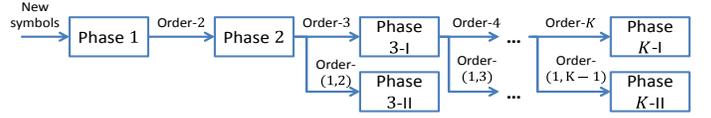}
\caption{Transmission flow}\label{fig:flow}
\end{figure}
\subsubsection{Transmission and Decoding Flow}
Similar to \cite{Abdoli13}, the $K$-phase transmission is illustrated in Figure \ref{fig:flow}. All the private symbols are transmitted in phase 1, generating order-2 symbols. Then, all the order-2 symbols are delivered in phase 2. At the end of phase 2, two types of higher order symbols are generated, namely order-3 and order-(1,2) symbols, which will be delivered in phase 3-I and 3-II respectively. This transmission is repeated till phase $K$, where order-$K$ and order-$(1{,}K{-}1)$ symbols are delivered.

A backward decoding is carried out. Specifically, each receiver recovers order-$K$ and $(1{,}K{-}1)$ symbols first. Then with their knowledge, order-$(K{-}1)$ symbols can be decoded. Repeatedly, order-$m$ symbols ($m{\geq}2$) are recovered using order-$(m{+}1)$ and $(1{,}m)$ symbols. At last, all the private symbols are decoded with the knowledge of order-$2$ symbols.

Considering that $N_1$ private symbols are sent in $T_1$ slots in phase 1, generating $N_2$ order-$2$ symbols, the sum DoF performance is expressed as $d_s(K){=}\frac{N_1}{T_1{+}\frac{N_2}{DoF_2(K)}}$, where $DoF_2(K)$ represents the DoF of sending order-$2$ symbols. Following the aforementioned transmission flow, $d_s(K)$ can be computed recursively as, for $2{\leq}m{\leq}K{-}1$,
\begin{IEEEeqnarray}{rcl}
DoF_m(K)&{=}&\frac{N_m}{T_m{+}\frac{N_{m{+}1}}{DoF_{m{+}1}(K)}{+}\frac{N_{1{,}m}}{DoF_{1{,}m}(K)}},\IEEEyessubnumber\label{eq:dofm}\\
DoF_K(K)&{=}&1,\IEEEyessubnumber\label{eq:DoFK}\\
DoF_{1{,}m}(K)&{=}&m{+}1,\IEEEyessubnumber\label{eq:DoF1m}
\end{IEEEeqnarray}
where $N_m$ and $N_{1{,}m}$ represent the number of order-$m$ and $(1{,}m)$ symbols respectively. $DoF_m(K)$ and $DoF_{1{,}m}(K)$ stand for the DoF of sending the corresponding symbols. $T_m$ refers to the number of slots in phase $m$-I. \eqref{eq:DoFK} is due to the fact that order-$K$ symbols are intended for all users and each user is equipped with a single antenna. \eqref{eq:DoF1m} will be shown in Section \ref{sec:phasem}. Next, the work is reduced to quantify $N_1$, $N_m$, $N_{m{+}1}$, $T_m$ and $N_{1{,}m}$ for $2{\leq}m{\leq}K{-}1$.

\subsubsection{Phase 1}\label{sec:phase1}
We consider a transmission mechanism with transmitter scheduling. Specifically, in a certain slot, a subset of $\mathcal{S}_n$ ($n{\leq}K$ to be shown later on) transmitters are active while others keep silent. Each of them delivers $n$ new symbols to the corresponding Rx. The received signal writes as
\begin{equation}
y_k{=}\mathbf{h}_{kk}^H\mathbf{W}_k\mathbf{x}_k{+}
\sum_{j{\in}\mathcal{S}_n{\setminus}k}\underbrace{\mathbf{h}_{kj}^H\mathbf{W}_j\mathbf{x}_j}_{u[j|k{,}j]},k{\in}\mathcal{S}_n,\label{eq:yphase1}
\end{equation}
where $\mathbf{W}_k$ is a full rank precoding matrix of size $K{\times}n$, while $\mathbf{x}_k$, of size $n{\times}1$, represent the symbol vector intended for Rx$k$. The second term refers to the overheard interferences. Each of them is an order-$2$ symbols as it is useful for Rx$k$ for interference cancelation/alignment, and for Rx$j$ as a useful side information.

It is straightforward that $n(n{-}1)$ order-$2$ symbols (e.g. $n$ receivers and each with $n{-}1$ interferers) are generated in a certain slot. Besides, since there are ${{K}\choose{n}}$ possible choices of $\mathcal{S}_n$, the similar transmission is repeated ${{K}\choose{n}}$ times for transmitter scheduling. Thus, one has
\begin{equation}
N_1{=}n^2{{K}\choose{n}},\quad
T_1{=}{{K}\choose{n}},\quad
N_2{=}n(n{-}1){{K}\choose{n}}.\label{eq:NT1}
\end{equation}
$d_s(K)$ is rewritten as
\begin{equation}
d_s(K){=}\frac{n^2}{1{+}\frac{n(n{-}1)}{DoF_2(K)}}.\label{eq:ds2}
\end{equation}
Hence, the optimal $n$ is chosen such that,
\begin{equation}
n^*{=}arg\max_{n{=}2{,}\cdots{,}K}\frac{n^2}{1{+}\frac{n(n{-}1)}{DoF_2(K)}}.\label{eq:nphase1}
\end{equation}
By evaluating the first and second order derivatives of \eqref{eq:ds2}, one can easily find that the global minimizer is given by $2DoF_2(K)$. As $n^*$ is an integer, we choose $n^*$ to be either $\lfloor2DoF_2(K)\rfloor$ or $\lceil2DoF_2(K)\rceil$. This leads to the maximization operator in \eqref{eq:ds0}. The remaining work is to find $DoF_2(K)$.
\begin{myremark}\label{rmk:n}
The new symbol transmission phase suggests that when there are $n$ active Tx, each delivering $n$ symbols to the corresponding Rx, it generates $n(n{-}1)$ order-$2$ symbols. Choosing $n{=}K$ fully makes use of the linear independent observations across the users, but it leads to a large number of order-$2$ symbols. On the other hand, a too small $n$ will lead to a waste of the linear independent observations across the users. Hence, $n^*$ highlights a trade-off between the utilization of the linear independent side information and the ability of delivering order-$2$ symbols.
\end{myremark}

\subsubsection{Phase $m$-I ($2{\leq}m{\leq}K$)}\label{sec:phasem}
We perform a $1$ Tx/$m$ Rx scheduling and employ the same higher order symbol transmission as in MAT. To be specific, in a certain slot and for a subset $\mathcal{S}_m$ of users, only one transmitter, i.e.Tx$k{,}k{\in}\mathcal{S}_m$, is active, delivering $K{-}m{+}1$ order-$m$ symbols, namely $\mathbf{u}[k|\mathcal{S}_m]$. We can write the transmitted signal and received signal as
\begin{IEEEeqnarray}{rcl}
\!\!\!\!\!\!\!\!\!\mathbf{s}_k&{=}&\mathbf{W}_k\mathbf{u}[k|\mathcal{S}_m]
{=}\mathbf{W}_k\left[u_1[k|\mathcal{S}_m]{,}{\cdots}{,}u_{K{-}m{+}1}[k|\mathcal{S}_m]\right]^T\!\!\!,\IEEEyessubnumber\label{eq:sphasem}\\
\!\!\!\!\!\!\!\!\!
y_l&{=}&\mathbf{h}_{lk}^H\mathbf{W}_k\mathbf{u}[k|\mathcal{S}_m]{,}{\forall}l{\in}\mathcal{S}_m,\IEEEyessubnumber\label{eq:ylphasem}\\
\!\!\!\!\!\!\!\!\!y_j&{=}&\mathbf{h}_{jk}^H\mathbf{W}_k\mathbf{u}[k|\mathcal{S}_m]
{=}u[k|\mathcal{S}_m{;}j]{,}{\forall}j{\notin}\mathcal{S}_m,\IEEEyessubnumber\label{eq:yjphasem}
\end{IEEEeqnarray}
where $\mathbf{W}_k$ is a full rank $K{\times}(K{-}m{+}1)$ matrix. As in \eqref{eq:yjphasem}, Rx$j{,}{\forall}j{\notin}\mathcal{S}_m$ overhears them as $u[k|\mathcal{S}_m;j]$. Indeed, there are $K{-}m$ different overheard interferences for a certain $k$ and $\mathcal{S}_m$. These $K{-}m$ side information are linearly independent of each other, and also linearly independent of $y_l{,}l{\in}\mathcal{S}_m$. Clearly, Rx$l{,}{\forall}l{\in}\mathcal{S}_m$ will be able to decode $\mathbf{u}[k|\mathcal{S}_m]$ if it obtains those $K{-}m$ side information $u[k|\mathcal{S}_m;j]{,}{\forall}j{\notin}\mathcal{S}_m$.

As there are $m$ choices of $k$ in $\mathcal{S}_m$ and there are ${{K}\choose{m}}$ possible choices of $\mathcal{S}_m$, we have
\begin{equation}
T_m{=}m{{K}\choose{m}},\quad
N_m{=}(K{-}m{+}1)T_m.\label{eq:NTm}
\end{equation}

To deliver the aforementioned side information, we \emph{generate order-$(m{+}1)$ symbols}. At the end of phase $m$-I (or beginning of phase $m{+}1$-I), for a fixed subset $\mathcal{S}_{m{+}1}$ of users and a fixed Tx$k{\in}\mathcal{S}_{m{+}1}$, an order-$(m{+}1)$ symbol is generated as
\begin{equation}
u[k|\mathcal{S}_{m{+}1}]{=}LC\left(u[k|\mathcal{S}_{m{+}1}{\setminus}j^\prime;j^\prime]{,}
{\forall}j^\prime{\in}\mathcal{S}_{m{+}1}{\setminus}k\right).\label{eq:ordermplus1}
\end{equation}
Clearly, $u[k|\mathcal{S}_{m{+}1}]$ is made up of $m$ different symbols as there are $m$ different Rx$j^\prime{,}j^\prime{\in}\mathcal{S}_{m{+}1}{\setminus}k$. Moreover, since each Rx$j^\prime$ has the knowledge of $u[k|\mathcal{S}_{m{+}1}{\setminus}j^\prime;j^\prime]$ and wishes to obtain the remaining side information, we need $m{-}1$ linear independent order-$(m{+}1)$ symbols for a fixed $\mathcal{S}_{m{+}1}$ and $k{\in}\mathcal{S}_{m{+}1}$. Thus, the total number of order-$(m{+}1)$ symbols is
\begin{equation}
N_{m{+}1}{=}(m{-}1)(m{+}1){{K}\choose{m{+}1}},\label{eq:Nmplus1}
\end{equation}
because there are ${{K}\choose{m{+}1}}$ choices of $\mathcal{S}_{m{+}1}$ and $m{+}1$ different transmitters in each $\mathcal{S}_{m{+}1}$.

\subsubsection{Phase $m$-II ($3{\leq}m{\leq}K$)}
For a fixed $\mathcal{S}_{m{+}1}$ and $k{\in}\mathcal{S}_{m{+}1}$, generating $m{-}1$ order-$(m{+}1)$ symbols does not guarantee the decodability of $u[k|\mathcal{S}_{m{+}1}{\setminus}j^\prime;j^\prime]{,}{\forall}j^\prime{\in}\mathcal{S}_{m{+}1}{\setminus}k$ at Rx$k$, as Rx$k$ requires $m$ independent $LC$ of them. Hence, one more $LC$ is needed. Then we generate an order-$(1{,}m)$ symbol as
\begin{equation}
u[k|k;\mathcal{S}_{m{+}1}{\setminus}k]{=}LC\left(u[k|\mathcal{S}_{m{+}1}{\setminus}j^\prime;j^\prime]{,}
{\forall}j^\prime{\in}\mathcal{S}_{m{+}1}{\setminus}k\right).\!\!\!\!\label{eq:order1m}
\end{equation}
The total number of order-$(1{,}m)$ symbols is
\begin{equation}
N_{1{,}m}{=}(m{+}1){{K}\choose{m{+}1}}.\label{eq:N1m}
\end{equation}

Note that after phase $m$-I, for a certain $\mathcal{S}_{m{+}1}$ and $k{\in}\mathcal{S}_{m{+}1}$, all symbols contained in $u[k|\mathcal{S}_{m{+}1}]$ will be available at Rx$j^\prime$, $j^\prime{\in}\mathcal{S}_{m{+}1}{\setminus}k$, thus $u[k|k;\mathcal{S}_{m{+}1}{\setminus}k]$ is only desired by Rx$k$ and will be aligned with $u[k|\mathcal{S}_{m{+}1}]$ at Rx$j^\prime$. Thus, the $m{+}1$ order-$(1{,}m)$ symbols can be transmitted simultaneously from the $m{+}1$ transmitters in phase $m$-II, leading to \eqref{eq:DoF1m}.

\begin{myremark}\label{rmk:K}
In phase $m$, since $K{-}m{+}1$ different order-$m$ symbols are delivered by one Tx per slot, the transmission in fact can be done using $K{-}m{+}1$ antennas. Specifically, we need $K{-}1$ antennas in phase $2$ while only a single antenna in phase $K$. Hence, if \eqref{eq:ds2} is maximized with $n^*{\leq}K{-}1$, we can conclude that the above transmission is applicable to the case where each Tx is equipped with $K{-}1$ antennas.
\end{myremark}

Plugging \eqref{eq:NTm}, \eqref{eq:Nmplus1} and \eqref{eq:N1m} into \eqref{eq:dofm}, one has
\begin{equation}
DoF_m(K){=}\frac{m(K{-}m{+}1)}{m{+}\frac{K{-}m}{m{+}1}{+}\frac{(m{-}1)(K{-}m)}{DoF_{m{+}1}(K)}}.\label{eq:dofm2}
\end{equation}
With \eqref{eq:DoFK} and \eqref{eq:DoF1m}, one has $DoF_2(K){=}\frac{1}{1{-}A_2(K)}$, where
\begin{equation}
A_2(K){=}\frac{1}{K{-}1}\sum_{l{=}2}^{K{-}1}\frac{K{-}l}{(l{-}1)(l{+}1)}.\label{eq:A2}
\end{equation}
The derivation is presented in the Appendix B. Combining with the optimization problem \eqref{eq:ds2} leads to the main theorem.

\section{Conclusion}\label{sec:conclusion}

This paper considers a $K$-user MISO IC with perfect delayed CSIT, where each transmitter is equipped with $K$ antennas. We propose a new scheme that achieves a greater sum DoF performance than the previously known results. The transmission is carried out with a $K$-phase process, which integrates the previously developed techniques, namely distributed higher order symbol generation, higher order symbol transmission in the MAT scheme. We also find that, the number of active transmitters in the new symbol transmission phase, namely $n^*$, should be properly chosen to balance the trade-off between the usage of linear independent side information and the ability of order-$2$ symbol transmission. The scheme is also applicable to the case where the transmitter is equipped with $K{-}1$ antennas if $n{\leq}K{-}1$. The case with general antenna configuration will be the future work.

\section*{Appendix}
\subsection{Derivation of the bounded value $\frac{64}{15}$}
For convenience, we approximate $2\mathcal{O}(K){\approx}\mathcal{O}_{i^*}(K)$, where $i^*$ is the solution to \eqref{eq:ds0}. For $K{\to}\infty$, to show $d_s{\approx}\frac{64}{15}$, it suffices to prove $\mathcal{O}(K){\approx}4$, namely $A_2(K){\approx}\frac{3}{4}$. From \eqref{eq:A2}, we have
\begin{align}
A_2(K){=}&\frac{K}{(K{-}1)}\sum_{l{=}2}^{K{-}1}\frac{1}{(l{-}1)(l{+}1)}{-}\frac{1}{K{-}1}\sum_{l{=}2}^{K{-}1}\frac{l}{(l{-}1)(l{+}1)}\nonumber\\
\stackrel{K{\to}{\infty}}{\approx}&\frac{K}{2(K{-}1)}\sum_{l{=}2}^{K{-}1}\frac{1}{l{-}1}{-}\frac{1}{l{+}1}\\
{=}&\frac{K}{2(K{-}1)}(1{+}\frac{1}{2}{-}\frac{1}{K{-}1}{-}\frac{1}{K}){\approx}\frac{3}{4}.
\end{align}

\subsection{Derivation of \eqref{eq:A2}}
Letting $A_m{\triangleq}1{-}\frac{1}{DoF_m}$, from \eqref{eq:dofm2}, for $2{\leq}m{\leq}K{-}1$, one has
\begin{equation}
A_m{=}\underbrace{\frac{(K{-}m)(m{-}1)}{m(K{-}m{+}1)}}_{B_m}A_{m{+}1}{+}
\underbrace{\frac{K{-}m}{(K{-}m{+}1)(m{+}1)}}_{C_m},\label{eq:Am}
\end{equation}
followed by
\begin{align}
B_mA_{m{+}1}{=}&B_mB_{m{+}1}A_{m{+}2}{+}B_mC_{m{+}1},\nonumber\\
B_mB_{m{+}1}A_{m{+}2}{=}&B_mB_{m{+}1}B_{m{+}2}A_{m{+}3}{+}B_mB_{m{+}1}C_{m{+}2},\nonumber\\
A_{K{-}1}\Pi_{i{=}m}^{K{-}2}B_i{=}&A_K\Pi_{i{=}m}^{K{-}1}B_i{+}C_{K{-}1}\Pi_{i{=}m}^{K{-}2}B_i,\nonumber
\end{align}
resulting in
\begin{equation}
A_m{=}A_K\Pi_{i{=}m}^{K{-}1}B_i{+}\sum_{l{=}m}^{K{-}1}C_l\Pi_{i{=}m}^{l{-}1}B_i.\label{eq:Am2}
\end{equation}
By the definition of $B_i$ and $C_i$ \eqref{eq:Am}, it is easily verified that
\begin{IEEEeqnarray}{rcl}
\Pi_{i{=}m}^{K{-}1}B_i&{=}&\frac{m{-}1}{(K{-}1)(K{-}m{+}1)},\IEEEyessubnumber\label{eq:sumB}\\
C_l\Pi_{i{=}m}^{l{-}1}B_i&{=}&\frac{m{-}1}{K{-}m{+}1}\frac{K{-}l}{l{+}1(l{-}1)}\IEEEyessubnumber\label{eq:sumC}
\end{IEEEeqnarray}
Substituting \eqref{eq:sumB} and \eqref{eq:sumC} into \eqref{eq:Am2}, and replacing $A_K{=}0$ and $m{=}2$, lead to \eqref{eq:A2}.

\bibliographystyle{IEEEtran}

\bibliography{Ref}

\end{document}